\documentclass{PoS}
 
\title{Eradication of singularities in the next-to-leading order RG evolution
for the $\Delta S=1$ effective Hamiltonian with 3 quark flavours}

\ShortTitle{Eradication of singularities in NLO RG evolution}

\author{\speaker{David H. Adams}\\
        Department of Physics and Astronomy, Seoul National University,
        Seoul, 151-747, South Korea\\
        E-mail: \email{dadams@phya.snu.ac.kr}}

\author{Weonjong Lee\\
        Frontier Physics Research Division and Center for Theoretical Physics,\\   
        Department of Physics and Astronomy, Seoul National University,
        Seoul, 151-747, South Korea\\
        E-mail: \email{wlee@phya.snu.ac.kr}}

\abstract{We consider the renormalization group (RG) evolution for the operators
in the $\Delta S=1$ effective Hamiltonian with 3 active quark flavors, which is
needed in the numerical analysis of data sets for $\e'/\e$
calculated in lattice QCD. 
Singularities are present in the original solution of Buras {\em et al.} at 
next-to-leading order (NLO). We show how these can be eradicated through a method
of analytic continuation to obtain the correct finite solution in this case.
Furthermore, we trace the origin of the singularities to a breakdown of the 
approach of Buras {\em et al.} in the 3 flavour case, and show how it can be 
rectified so that singularitites are absent from 
the beginning.}

\FullConference{The XXV International Symposium on Lattice Field Theory\\
		 July 30-4 August 2007\\
		 Regensburg, Germany}

\def\lb{\lbrack}
\def\rb{\rbrack}

\def\be{\begin{eqnarray}}
\def\ee{\end{eqnarray}}

\def\H{{\cal H}}
\def\e{\epsilon}

\begin{document}

\section{Introduction}

Weak $\Delta S=1$ decays of hadrons can be described in the framework of the
Standard Model by an effective Hamiltonian $\H_{eff}$ obtained by integrating
out the degrees of freedom associated with heavy quarks and gauge bosons
\cite{1,2,3,4,5}. It governs interesting aspects of kaon physics; in particular
the direct CP violation parameter
$\epsilon'/\epsilon$ and the $\Delta I=1/2$ rule. The effective Hamiltonian has
the form
\be
\H_{eff}=\frac{G_F}{\sqrt{2}}\sum_iC_i(\mu,g,e)Q_i(\mu,g,e)
\label{1.1}
\ee
where $C_i(\mu,g,e)$ denotes the Wilson coefficients, the $Q_i(\mu,g,e)$'s  
are a basis of renormalized 4-fermion operators contributing to the effective
Hamiltonian, $\mu$ is the scale, $g$ the strong coupling and $e$ the EM coupling. 
We take the basis operators $Q_i$ to be the ones specified by Buras {\em et al.}
in Section 2 of Ref.\cite{1}. There are 10 of these and their origins are as 
follows: $Q_{1-2}$ are from $W$-exchange, $Q_{3-6}$ are from QCD penguin diagrams
while $Q_{7-10}$ are from electroweak penguin diagrams.

The energy scale in kaon decays is about 500 MeV, so 
the hadronic matrix elements of $\H_{eff}$ are dominated by the strong interaction
and must be calculated non-perturbatively. The lattice approach is the only 
possibility for doing this at present, and is currently being used; see Ref.'s
\cite{7,8,9,10}.
The procedure for calculating the matrix elements from the lattice 
can be summarized as follows.
First one calculates the matrix elements of the appropriate lattice operators
$Q_j^{latt}(a)$. Next one matches to the continuum operators at some scale
$q^*\approx1/a$ in a continuum renormalization scheme, 
{\em e.g.}, Naive Dimensional Regularization (NDR):
\be
Q_i(q^*)=z_{ij}(q^*a)Q_j^{latt}(a)
\label{1.2}
\ee
The matching factors $z_{ij}(q^*a)$ can be calculated in perturbation 
theory \cite{11}; they involve $\log(q^*a)$ so the continuum scale $q^*$ must be 
close to $1/a$ to avoid large logs. Finally, one uses RG evolution to
run the continuum operators $Q_i(\mu)$ from the scale $\mu=q^*$ down to 
$\mu=m_c$ (the charm 
quark mass) where they can be combined with the known Wilson 
coefficients\footnote{The Wilson coefficients are all known up to NLO in 
peturbation theory \cite{2}.} at the scale $\mu=m_c$
to get the matrix elements of $\H_{eff}$.\footnote{Note that $\H_{eff}$ itself 
is independent of the scale, so 
we can combine the operators and Wilson coefficients at any chosen scale to 
determine its matrix elements.}

The RG evolution operator required in the final step above has been calculated
up to next-to-leading order (NLO) by Buras {\em et al.} \cite{1}. 
However, the expression obtained there contains singularities in the 
case where there are 3 active quark flavors. Therefore it cannot
be used for extracting the matrix elements from lattice QCD simulations with 
$N_f=2+1$ dynamical fermion flavors. This is a major problem since unquenched
lattice simulations with $2+1$ sea quark flavors are currently underway with a 
variety of fermion discretizations (see, {\em e.g.}, \cite{Lee(lat06)} and 
references therein) and will be addressing kaon physics such as 
$\epsilon'/\epsilon$ and the $\Delta I=1/2$ rule in the coming future.
It is therefore imperative to deal with the singularity problem in the NLO 
expression for the RG evolution operator. The full RG evolution operator
is known to be singularity-free, so the singularity in the NLO expression in the
3 flavor case should be a removable artifact. In this paper we review the
solution to this problem given recently in Ref.\cite{AL}. The singularity
is eradicated by a method of analytic continuation to obtain the correct finite 
NLO expression for the RG operator in the 3 flavor case.
Furthermore, we trace the origin of the singularities in the work of Buras 
{\em et al.} to a breakdown of their approach in the 3 flavor case, and show how 
it can be rectified so that singularities are absent 
from the beginning.

\section{Review of RG evolution and the singularity problem at NLO}

The running of the $Q_i(\mu)$'s and $C_i(\mu)$'s is governed by the RG evolution
equation resulting from $\frac{d}{d\mu}\H_{eff}=0$. Combining the operators
and Wilson coefficients into vectors $\vec{Q}$ and $\vec{C}$ respectively, 
the evolution is given by $10\times10$ matrices acting on these vectors. The 
evolution of $\vec{Q}(\mu)$ is obviously inverse to that of $\vec{C}(\mu)$
so it suffices to determine the latter, which is what we do in
the following. The running of the EM coupling $e$ is negligible over the range
of scales that are relevant for the present considerations, so we treat it
as constant in the following, as was done in Ref.\cite{1}. 

The RG equation for $\vec{C}(\mu)$ is
\be
\Big\lb\mu\frac{\partial}{\partial\mu}+\beta(g,e)\frac{\partial}{\partial g}\Big\rb
\vec{C}=\gamma^T(g,e)\vec{C}
\label{2.1}
\ee
where $\gamma(g,e)$ is the $10\times10$ anomalous dimension matrix (given below)
and $\beta(g,e)$ is the beta-function, given by
\be
\beta(g,e)=-\beta_0\frac{g^3}{16\pi^2}-\beta_1\frac{g^5}{(16\pi^2)^2}
-\beta_{1e}\frac{e^2g^3}{(16\pi^2)^2}+\dots
\label{2.2}
\ee
with
\be
\beta_0=11-\frac{2}{3}f\ ,\qquad\beta_1=102-\frac{38}{3}f\ ,
\qquad\beta_{1e}=-\frac{8}{9}(u+\frac{d}{4})
\label{2.3}
\ee
where $u$ and $d$ denote the number of active $u-$ and $d-$type flavors, 
respectively, and $f=u+d$ $(=N_f)$ is the total number of active flavors.

From the RG equation (\ref{2.1}) the running of $\vec{C}(\mu)$ is found to be
given by
\be
\vec{C}(m_1)=U(m_1,m_2)\vec{C}(m_2)
\label{2.4}
\ee
where the evolution matrix is 
\be
U(m_1,m_2)=T_g\exp\Big(\int_{g(m_2)}^{g(m_1)}dg'\,\frac{\gamma^T(g',e)}{\beta(g',e)}
\Big)
\label{2.5}
\ee
with the dependence $g=g(m)$ specified by $m\frac{dg}{dm}=\beta(g,e)$.
Here $T_g$ denotes $g-$ordering; it is required since generally
$[\gamma(g_1),\gamma(g_2)]\ne0$ for $g_1\ne g_2$. 

To evaluate $U(m_1,m_2)$ we need to know the anomalous dimension matrix 
$\gamma(g,e)$. It is determined from the renormalization constant matrix 
relating the bare and renormalized operators: $Q_i^{(0)}=Z_{ij}(\mu)Q_j(\mu)$
and
\be
\gamma(g,e)=Z^{-1}\frac{d}{d\log\mu}Z
\label{2.6}
\ee
This depends on $g$ and $e$ through $\alpha_s=\frac{g^2}{4\pi}$ and
$\alpha=\frac{e^2}{4\pi}$, and has been calculated perturbatively up to 2 loops 
in the NDR scheme, whereby the matrices in the following expansions have 
been determined (see \cite{1} and the references therein):
\be
\gamma(g,e)=\gamma_s(g^2)+\frac{\alpha}{4\pi}\Gamma(g^2)+O(\alpha^2)
\label{2.7}
\ee
where the pure QCD part is
\be
\gamma_s(g^2)=\frac{\alpha_s}{4\pi}\gamma_s^{(0)}
+\frac{\alpha_s^2}{(4\pi)^2}\gamma_s^{(1)}+\dots
\label{2.8}
\ee
and the leading order QED correction in (\ref{2.7}) is specified by
\be
\Gamma(g^2)=\gamma_e^{(0)}+\frac{\alpha_s}{4\pi}\gamma_{se}^{(1)}+\dots
\label{2.9}
\ee
For later use we note that
\be
\frac{\gamma(g,e)}{\beta(g,e)}\ &=&\ -\frac{4\pi}{\beta_0g^3}
\left\lb
\begin{array}{l}
\alpha_s\gamma_s^{(0)}+\frac{\alpha_s^2}{4\pi}(\gamma_s^{(1)}-
\frac{\beta_1}{\beta_0}\gamma_s^{(0)}) \\
+\ \alpha\Big(\gamma_e^{(0)}+\frac{\alpha_s}{4\pi}(\gamma_{se}^{(1)}-
\frac{\beta_1}{\beta_0}\gamma_e^{(0)}-\frac{\beta_{1e}}{\beta_0}\gamma_s^{(0)})
+O(\alpha_s^2)\Big)+O(\alpha^2) \\
\end{array}
\right\rb
\label{2.10} \\
&=&-\frac{\gamma_s^{(0)}}{\beta_0g}+O(g)+O(\alpha)
\label{2.11}
\ee
The expansion of the evolution operator in the EM coupling takes the form
\be
U(m_1,m_2)=U_s(m_1,m_2)+\frac{\alpha}{4\pi}R(m_1,m_2)+O(\alpha^2)
\label{2.12}
\ee
where the pure QCD evolution is
\be
U_s(m_1,m_2)=T_g\exp\Big(\int_{g(m_2)}^{g(m_1)}dg'\,\frac{\gamma_s^T(g')}{\beta_s(g')}
\Big)
\label{2.13}
\ee
and the leading additional contribution to the QCD evolution in the presence
of EM interactions in (\ref{2.12}) is given by (see \cite{1})
\be
R(m_1,m_2)=
\int_{g(m_2)}^{g(m_1)}dg'\,\frac{U_s(m_1,m')\Gamma^T(g')U_s(m',m_2)}{\beta_s(g')}
\label{2.14}
\ee
where $g'=g'(m')$. The EM contribution to the beta-function has been ignored
in (\ref{2.13})--(\ref{2.14}): $\beta_s(g)=\beta(g,0)$ 
so the expressions are valid when the 
$\beta_{1e}$ term in (\ref{2.2}) is dropped, which is a justifiable approximation 
made in Ref.\cite{1}. However, the generalization of the NLO expressions for the 
evolution matrix to the case where the $\beta_{1e}$ term is not dropped is
straightforward: In light of (\ref{2.10}) it can be obtained simply by
replacing $\gamma_{se}^{(1)}\to\gamma_{se}^{(1)}
-\frac{\beta_{1e}}{\beta_0}\gamma_s^{(0)}$ in the relevant expressions \cite{AL}.

In the remainder of this paper we restrict our attention to the pure QCD evolution
$U_s(m_1,m_2)$ which is where the aforementioned singularity problem arises at NLO
in the 3 flavor case. Once the singularity is eradicated, the new finite expression
needs to be used in the NLO evaluation of (\ref{2.14}) for $R(m_1,m_2)$. We
omit that part here, referring to Ref.\cite{AL} for the details and result.

We begin by recalling the leading logarithmic approximation to $U_s(m_1,m_2)$:
it is obtained by keeping
only the leading contribution to $\frac{\gamma_s^T(g')}{\beta_s(g')}$ in the 
integrand in (\ref{2.13}), which is shown in (\ref{2.11}). It is useful to 
diagonalize $\gamma_s^{(0)T}\,$:
\be
\gamma_D^{(0)}=V^{-1}\gamma_s^{(0)T}V \qquad\mbox{diagonal matrix}
\label{2.15}
\ee
then the leading logarithmic approximation is found to be
\be
U_s^{(0)}(m_1,m_2)=V\Big(\frac{\alpha_s(m_2)}{\alpha_s(m_1)}
\Big)^{\gamma_D^{(0)}/2\beta_0}V^{-1}
\label{2.16}
\ee 
The NLO contribution to the evolution matrix can now be found starting from the
ansatz
\be
U_s(m_1,m_2)=\Big(1+\frac{\alpha_s(m_1)}{4\pi}J\Big)U_s^{(0)}(m_1,m_2)
\Big(1-\frac{\alpha_s(m_2)}{4\pi}J\Big)
\label{2.17}
\ee
and using the RG equation for $U_s(m_1,m_2)$ (given by (\ref{2.1}) with
$\beta(g,e)$ and $\gamma(g,e)$ replaced by $\beta_s(g)$ and $\gamma_s(g)$, 
respectively) to derive an equation for $J$. From solving this equation at lowest
order in $\alpha_s$ Buras {\em et al.} find \cite{1,3}
\be
J=VSV^{-1}
\label{2.18}
\ee
where
\be
S_{ij}=\delta_{ij}\gamma_i^{(0)}\frac{\beta_1}{2\beta_0^2}
-\frac{G_{ij}}{2\beta_0(1+a_i-a_j)}
\label{2.19}
\ee
with
\be
G=V^{-1}\gamma_s^{(1)}V\quad,\qquad\mbox{$a_i=$ $i$'th diagonal
element of $\gamma_D^{(0)}/2\beta_0$}
\label{2.20}
\ee
From (\ref{2.19}) we see that the resulting NLO expression for the evolution
matrix has a singularity if $1+a_i-a_j=0$.
As mentioned in \cite{1}, this happens for $(i,j)=(8,7)$  
when there are 3 active quark flavors ($f=3$).

\section{Solution of the singularity problem via analytic continuation}

The expression (\ref{2.17}) can be written up to NLO as
\be
U_s^{NLO}(m_1,m_2)=U_s^{(0)}(m_1,m_2)+\frac{1}{4\pi}VA(m_1,m_2)V^{-1}
\label{3.1}
\ee
where
\be
VA(m_1,m_2)V^{-1}=\alpha_s(m_1)JU_s^{(0)}(m_1,m_2)
-\alpha_s(m_2)U_s^{(0)}(m_1,m_2)J\,.
\label{3.2}
\ee
Inserting the expression (\ref{2.16}) for $U_s^{(0)}(m_1,m_2)$ leads to
\be
A_{ij}(m_1,m_2)=S_{ij}\Big\lb\alpha_1\Big(\frac{\alpha_2}{\alpha_1}\Big)^{a_j}
-\alpha_2\Big(\frac{\alpha_2}{\alpha_1}\Big)^{a_i}\Big\rb
\label{3.3}
\ee
where $\alpha_k\,\equiv\,\alpha_s(m_k)$ for $k=1,2$.
In the singular case ($f=3$, $i=8$, $j=7$) we now regularize $S_{ij}$ by replacing
\be
a_j\to a_j+\e 
\label{3.4}
\ee
in (\ref{2.19}). The regularized quantity then becomes
$S_{ij}=\Big(\frac{G_{ij}}{2\beta_0}\Big)\frac{1}{\e}$. Inserting this into 
(\ref{3.3}), and making the same regularization (\ref{3.4}) there, we find
\be
A_{ij}(m_1,m_2)&=&\Big(\frac{G_{ij}}{2\beta_0}\Big)\frac{1}{\e}
\Big\lb\alpha_1\Big(\frac{\alpha_2}{\alpha_1}\Big)^{a_j+\e}
-\alpha_2\Big(\frac{\alpha_2}{\alpha_1}\Big)^{a_i}\Big\rb \nonumber \\ 
&=&\Big(\frac{G_{ij}}{2\beta_0}\Big)\frac{1}{\e}
\alpha_2\Big(\frac{\alpha_2}{\alpha_1}\Big)^{a_i}\Big\lb\e\log\Big(\frac{\alpha_2}
{\alpha_1}\Big)+O(\e^2)\Big\rb \nonumber \\
&\stackrel{\e\to0}{=}&\Big(\frac{G_{ij}}{2\beta_0}\Big)
\alpha_2\Big(\frac{\alpha_2}{\alpha_1}\Big)^{a_i}\log\Big(\frac{\alpha_2}{\alpha_1}
\Big) 
\label{3.5}
\ee
Thus a finite expression is obtained in the limit where the regularization is
lifted. Substituting this into (\ref{3.1}) we get a finite NLO expression for 
the evolution operator. This solution of the singularity problem clearly amounts to 
analytic continuation of the evolution operator (regarded as a 
function of the matrix $\gamma_s^{(0)}$).

\section{Understanding the ``singularity'' from first principles}

In the following it is convenient to use the notation $U_s(g_1,g_2)$ rather
than $U_s(m_1,m_2)$; this is justified since the dependence on $m_1,m_2$ 
enters exclusively through $g(m_1),g(m_2)$.

By general arguments the full evolution operator (for pure QCD) can be written as
\be
U_s(g_1,g_2)=\Big(1+\frac{g_1^2}{16\pi^2}J(g_1)\Big)U_s^{(0)}(g_1,g_2)
\Big(1+\frac{g_2^2}{16\pi^2}J(g_2)\Big)^{-1}
\label{4.1}
\ee
Using the RG equation
\be
\frac{d}{dg}U_s(g,g_0)=\frac{\gamma_s^T(g)}{\beta_s(g)}U_s(g,g_0)
\label{4.2}
\ee
to derive an equation for $J(g)$ at leading order, one finds
{\em different} equations in the 
``singular'' case $1+a_i-a_j=0$ and ``non-singular'' case \cite{AL}.

In the non-singular case, taking $J(g)=J+O(g)$ leads to a consistent solution
for the constant matrix $J$. This is the solution found by Buras {\em et al.}
that we reviewed in \S2. On the other hand, in the singular case, the ansatz
$J(g)=J+O(g)$ does {\em not} admit a consistent solution. In fact, it turns out
that $S_{ij}(g)=(V^{-1}J(g)V)_{ij}$ must {\em diverge} for $g\to0$. Specifically,
in the singular case $1+a_i-a_j=0$ we found in Ref.\cite{AL} that the 
leading order equation for $S_{ij}(g)$ is as follows:
\be
\beta_0\,g\frac{d}{dg}S_{ij}(g)=-G_{ij}
\label{4.3}
\ee
The solution is 
\be
S_{ij}(g)=-\frac{G_{ij}}{\beta_0}\log(g)+c_{ij}
=-\frac{G_{ij}}{2\beta_0}\log(\alpha_s)+c_{ij}'
\label{4.4}
\ee
where $c_{ij}$ is an undetermined integration constant and 
$c_{ij}'=c_{ij}-\frac{G_{ij}}{2\beta_0}\log(4\pi)$. Thus $S_{ij}(g)$ diverges
for $g\to0$ as claimed. Note however that $g^2S_{ij}(g)$ vanishes for $g\to0$,
implying that $g^2J(g)$ vanishes in this limit as it should in order for 
(\ref{4.1}) to reduce to $U_s^{(0)}(m_1,m_2)$ in the small $g$ limit.
We also note that $S_{ij}(g)$ is actually
a function of $g^2$ (or $\alpha_s$) as it should be.
Finally, our previous expression (\ref{3.5}) for 
$A_{ij}(m_1,m_2)$ is readily reproduced from (\ref{4.4}) 
(see \cite{AL}),\footnote{The occurrences of the undetermined constant $c_{ij}'$ 
from (\ref{4.4}) are found to cancel out in $A_{ij}(m_1,m_2)$
\cite{AL}.} so the first principles solution of the singularity problem presented 
here agrees as it should with the finite NLO result for the evolution
operator obtained via analytic continuation in the previous section.

\section{Conclusions}

We have eradicated the singularities in the original solution of Buras {\em et al.}
to get a finite expression for the RG evolution matrix at NLO in the 3 flavor case.
This is essential for being able to evaluate the matrix elements of the 
$\Delta S=1$ effective Hamiltonian from the lattice with $2+1$ sea quark flavors.

The breakdown of the ansatz $J(g)=J+O(g)$ in Buras {\em et al.}'s approach,
and the rectification discussed here, illustrate some general subtleties to bear in 
mind when evaluating evolution operators in general.

\section{Acknowledgements}

This research is supported by the KICOS international cooperative
research program (KICOS grant K20711000014-07A0100-01410), 
by the KRF grant KRF-2006-312-C00497, by the BK21 program of 
Seoul National University, and by the DOE SciDAC-2 program.

\end{document}